\documentclass[a4paper]{jpconf}
\usepackage{graphicx}
\usepackage{amsfonts}
\usepackage{amsmath}
\usepackage{amssymb}
\usepackage{myaasmacros}

\begin{document}
\title{Looking for Hall attractor in astrophysical sources}

\author{S.B. Popov$^{1}$, A.P. Igoshev$^{2}$, R. Taverna$^{3}$, R.
Turolla$^{3}$}

\address{$^{1}$ Sternberg Astronomical Institute, Lomonosov Moscow State
University, Russia\\
$^{2}$ Department of Astronomy/IMAPP Radboud University Nijmigen,
Netherlands\\
$^{3}$ Department of Physics and Astronomy, University of Padua, Italy}

\ead{polar@sai.msu.ru}

\begin{abstract}
 Recently, numerical calculations of the magnetic field evolution in neutron stars demonstrated the possible existence of a Hall attractor, a stage at which the evolution of the field driven by the Hall cascade ends. The existence of such a stage in neutron star magnetic evolution is very important, and can be potentially probed by observations. Here we discuss three types of objects which could have reached this stage. First, we briefly describe the evolution of normal radio pulsars with ages about a few hundred thousand years. Then we analyse in more detail observations of RX J1856.5-3754, one of the Magnificent Seven, focusing on the surface temperature distribution and comparing model predictions with the temperature map inferred from X-ray observations. Finally, we discuss the necessity of the Hall attractor stage to explain the hypothetical existence of accreting magnetars. We conclude that at the moment there is no direct confirmation of the Hall attractor stage in known sources. However, more detailed observations in the near future can demonstrate existence (or absence) of this stage of the crustal magnetic field evolution.
\end{abstract}

\section{Introduction}

Magnetic field evolution is one of the most important features  in the biography of a neutron star (NS). The most natural behavior of the field is to decay, due to the progerssive dissipation of the supporting electric currents (for a review, see e.g. \cite{2009ASSL..357..319G}). However, in some instances the external field can increase due to the re-emergence of the surface field buried by an accretion episode (see recent calculations in \cite{2016MNRAS.462.3689I}). 

The evolution of  the crustal component of the NS magnetic field occurs on two distinct time-scales: $\tau_\mathrm{Ohm}$, related  to  Ohmic resistivity (i.e. to a finite conductivity of the crust) and $\tau_\mathrm{Hall}$, related to the Hall cascade. The Hall evolution is in principle non-dissipative. Nevertheless, it redistributes the magnetic energy from the larger spatial scales (e.g that of the dipole field) to the smaller ones (those of higher order multipoles) over which Ohmic decay acts faster. Ultimately, this  causes the decay of the dipole component and an enhanced release of magnetic energy.

The timescale of the Hall evolution is
\begin{equation}
\tau_\mathrm{Hall} = \frac{4\pi e n_\mathrm{e} L^2}{cB(t)},
\label{e:hall_t}
\end{equation}
with $n_\mathrm{e}$ the  electron density, $e$ the elementary charge, $B(t)$ the magnetic field, $L$ the typical spatial scale of electric currents and $c$ the speed of light \cite{2004ApJ...609..999C}.

The Hall cascade stops when (and if) the so-called {\it Hall attractor} is reached. The existence of this stage was proposed by \cite{2014PhRvL.112q1101G,2014MNRAS.438.1618G} and then independently confirmed by \cite{2015PhRvL.114s1101W} (see also \cite{2014ApJ...796...94M}, where the authors studied stability of Hall equilibria in NS crust). In \cite{2014PhRvL.112q1101G} it was shown that this stage is reached after a few initial Hall time-scales. Relaxation of the Hall process after early very active phase was also noted before by \cite{2009A&A...496..207P}.
Recently, the attractor stage was confirmed by \cite{2017arXiv170909167B} in simulations which include the core magnetic field evolution.

Up to now there are no direct observational evidences in favour of the Hall attractor stage. In this note we discuss three types of sources, and outline three different strategies to identify the presence of the Hall attractor. 

\section{Hall attractor and normal radio pulsars}
The large number of known normal radio pulsars gives an opportunity to probe the magneto-rotational evolution of NSs at least on time scales $\lesssim 10$~Myr (for older sources selection effects become very important). However, till present, population synthesis or/and studies of individual objects of this type did not provide a clear evidence for magnetic field decay for NSs with initial field $\sim10^{12}$~--~$10^{13}$~G during the radio pulsar phase.

\begin{figure}
\includegraphics[width=1\linewidth]{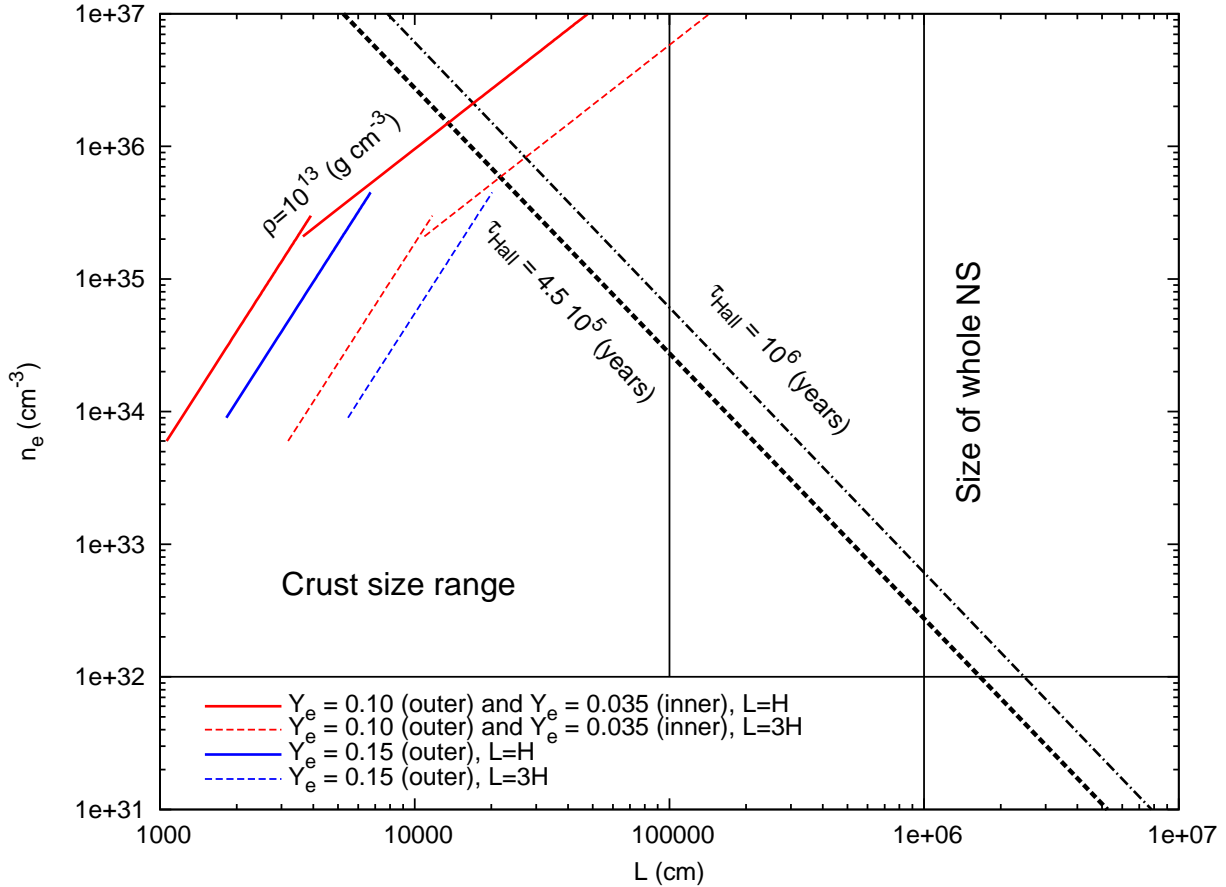}
\caption{
Spatial scale and electron density. Black thick lines show the locus where 
 the Hall timescale is
$\tau_\mathrm{Hall}=4.5\times 10^5$ yr (dotted line) and $\tau_\mathrm{Hall}= 10^6$ years
(dot-and-dashed line). The red thick lines correspond to taking the spatial scale  
equal to the local pressure scale height in the crust. 
The break corresponds to the density $\rho=10^{13}$ g cm$^{-3}$. 
The thinner dotted red line corresponds to the same relation but 
when the spatial scale is  three times larger than the local pressure scale height.
}
\label{contour}
\end{figure}

In \cite{2014MNRAS.444.1066I} we applied a modified pulsar current method to study the magnetic field evolution of radio pulsars. It was shown that, over several hundred thousand years, the fields of radio pulsars decay by a factor two with a time-scale $\sim 4\times 10^5$~yrs. However, for ages $\gtrsim10^6$~yrs no significant decay is seen. To identify the cause responsible for this decay we plot in Fig. \ref{contour} the electron density $n_\mathrm{e}$ against the characteristic spatial scale of currents $L$. The Hall time-scale with the value $4\times 10^5$~yrs corresponds to the black dashed line, see eq. (\ref{e:hall_t}).  In the crust of a real NS each electron density corresponds to a unique spatial scale due to the pressure gradient \cite{2004ApJ...609..999C}. This relation is shown with red and blue lines depending on the exact relation between the pressure height gradient $H$ and $L$ (dashed lines corresponds to $L$ = 3$H$), and the electron fraction $Y_\mathrm{e}$. 

Our region of interest is
at the intersection of the black dashed line and one of the colored lines.
From Fig. \ref{contour} one can see that the Hall timescale $\sim 4 \times
10^5$~yrs corresponds to the case
in which electric currents are mostly situated deep in the crust 
at electron densities about $(0.5$~--~$2.0)\times 10^{36}$ cm$^{-3}$,
that corresponds to a mass density about $(2.4$~--~$ 9.5) \times 10^{13}$~g~cm$^{-3}$. This is a realistic range of densities. However, the onset of the Hall attractor in this case  occurs at an age $\sim 2$~Myrs, which is  too large to fit the data. 
Thus, a fine-tuning of the parameters is required in order to match the predictions of the Hall cascade and attractor model with the findings from \cite{2014MNRAS.444.1066I} (see \cite{2015AN....336..831I}).

A different possibility to explain time-scales $\sim 4 \times 10^5$~yrs is related to the Ohmic decay due to phonons. The desired value is obtained for quite realistic conditions. Also, this type of decay stops naturally after $\sim 1$ Myr due to  cooling  when the NS temperature drops below the critical value, $T_\mathrm{U}$. 



Therefore, we think that the second possibility is more
probable. It may be that the Hall effect also contributes to the field decay in
normal radio pulsars, but the main contribution,
according to our analysis \cite{2015AN....336..831I}, is due to Ohmic decay on phonons. Thus, there are no clear signs of the Hall attractor in the case of normal radio pulsars.

\section{Hall attractor and the Magnificent Seven}     

Modern scenarios of magnetic field evolution with the Hall attractor can make important, and potentially testable, predictions about NS properties.
Calculations made under the assumption of axial symmetry by \cite{2014MNRAS.438.1618G, 2014PhRvL.112q1101G}  have shown that Hall evolution saturates after a few $\tau_\mathrm{Hall}$. The  field then reaches some stable configuration, and the successive evolution is driven mainly by the relatively slow Ohmic dissipation. The stage when the Hall cascade stops -- the Hall attractor -- is reached in $\lesssim 1$ Myr for magnetar-like initial fields. This picture was confirmed through 3D numerical simulations by \cite{2015PhRvL.114s1101W}, and recently generalized by \cite{2017arXiv170909167B} (however,  different time-scales were obtained in this investigation). The Hall cascade moves the crustal field towards the crust-core boundary, where it dissipates \cite{2014MNRAS.438.1618G}, so that the field looks more core-centered. According to \cite{2014MNRAS.438.1618G} the Hall attractor has a well-defined property: when the field structure is stabilized, its poloidal part mainly consists of dipole and octupole components
(with a small $l=5$ multipole in addition).
Since the surface thermal distribution in cooling, magnetized INSs is determined by the structure of their crustal field, the analysis of their spectral properties can provide a direct test for the Hall attractor scenario. We used this approach in \cite{2017MNRAS.464.4390P}.

In order to compare the predictions on the field evolution \cite{2014MNRAS.438.1618G, 2014PhRvL.112q1101G} with observations, we have to select a sample of INSs which are presumably close enough to the Hall attractor stage. In particular, these sources need to be of the right age ($\lesssim 1$ Myr) and possess initially high fields;  moreover, the emission from their cooling surface should be not polluted by strong contributions of non-thermal magnetospheric emission. The most promising candidates belong to a small group of near-by cooling isolated NSs dubbed the Magnificent Seven (M7). They are characterized by stable, purely thermal emission with temperatures $\sim 50$-$100$~eV. Typically, their spectra can be fitted by one or two blackbody components with the addition of a broad absorption feature at few hundred eVs. The pulsed fraction is usually low ($\lesssim 20$\%) and their luminosities are in the range
$L\sim10^{31}$--$10^{32}$~erg~s$^{-1}$ (see a review in \cite{2009ASSL..357..141T}).  

Among the Magnificent Seven RX J1856.5-3754 looks particularly promising, since, in the first place, it can be closer to the Hall attractor stage, having a lower blackbody temperature ($\sim 50$--$60$ eV), and hence luminosity, than other sources. Furthermore, its dipolar B-field, $\sim 10^{13}\, \mathrm G$, is one of the weakest among the Seven. 
In addition, this source is very well studied, e.g. in comparison with RX J0420.0-5022, which also seems to be close to the attractor stage. RX J1856.5-3754 is the prototype and the brightest member of the class, and its spectral properties are very well-characterized: the pulsed fraction is the lowest, $\sim 1$\%, no variability was detected so far and the X-ray spectrum is well fitted by two blackbody components with $kT^\infty_1 \sim 61$--$62$~eV, $R^\infty_1\sim 4.5$--$5$~km and $kT^\infty_2 \sim 39$~eV, $R^\infty_2\sim 11$--$16$~km (see \cite{2012A&A...541A..66S}). 

Details of the methods we used are described in \cite{2017MNRAS.464.4390P}. Here we present the main results. 
We found that, at the Hall attractor stage and fitting the multi-temperature spectrum with two blackbodies, the area corresponding to the hard component is larger than that related to the soft one, contrary to what observations of RX J1856.5-3754 show. We conclude that in the case of RX J1856.5-3754 (and, most probably, also in most or all other M7 sources) the Hall attractor is not reached, or the field structure at this stage is different from what current models predict.

\section{Hall attractor and accreting magnetars}  

Many NSs have been observed in X-ray binaries. In many cases, the magnetic field of the compact object can be measured via cyclotron line observations, or estimated by modeling the spin behavior or/and the properties of the accretion flow. All reliable measurements provide values $\sim 10^{12}$~--~$10^{13}$~G. Indirect estimates result in a wider range.

Among NS X-ray binaries of special interest for our discussion are those containing a so-called accreting magnetar. For these NSs  magnetic field strenghts $\sim10^{14}$~G have been inferred \cite{2010A&A...515A..10D, 2012MNRAS.425..595R, 2012ApJ...757..171F,2014MNRAS.437.3664H}, although no direct measurements are available so far. The existence of such systems can potentially contradict the standard scenario for NS magnetic field evolution, according to which magnetar-scale fields are typically dissipated on time-scales $\lesssim 10^6$~yrs, shorter than the inferred age of the neutron star. 
In \cite{2017arXiv170910385I} we explored different possibilities to form middle-aged magnetars varying the parameters which control the magnetic field evolution.

We constructed a model of the magnetic field evolution which includes the Hall cascade, Ohmic dissipation due to phonons and impurities, and the Hall attractor. Varying the model parameters, we tried to fit the properties of two accreting magnetar candidates: ULX M82 X-2 and 4U 0114+65. For the latter one we also presented a new estimate of its age based on backward integration of its trajectory in the Galactic potential within the Bayesian statistics. 


The bottom line of this study is the following: to form an accreting magnetar with an age $\gtrsim$~few Myrs it is necessary to include three main ingredients: (i) the field should have reached the Hall attractor stage; (ii) scatterings on phonons should become negligible after $\sim$few hundred thousand years (i.e. a relatively rapid cooling to the critical temperature when phonons are not important anymore should occur); and (iii) impurities have to be negligible (i.e. the parameter $Q$ which characterizes the role of impurities should assume low values, $\lesssim$~a few). Notice that the latter option (low $Q$) is considered to be untypical for magnetars, see \cite{2013NatPh...9..431P}. 
In addition, new approaches to accretion physics allow to explain parameters of many proposed magnetar candidates with standard fields $\lesssim$~few~$10^{13}$~G  (see \cite{2014EPJWC..6402002P} and references therein). 
Multiple contradicting magnetic field estimates prevent us from concluding that the Hall attractor definitely exists in the accreting magnetar candidates. 

\section{Discussion and conclusions}       


Typically, when we are dealing with high mass X-ray binaries, we do not know the precise age of the system. This complicates the comparison between model predictions and observations. However, there are several important exceptions, such as the source SXP 1062 observed in a supernova remnant (see recent data and references in \cite{2017arXiv170605002G}). 
This object has been already proposed as an accreting magnetar candidate  \cite{2012ApJ...757..171F}. On the other hand, its properties can be explained in the framework of a decaying field model \cite{2012MNRAS.421L.127P}, with the present day value of $B$ below the magnetar scale. The existence of different accretion models does not allow at the moment to make a definite estimate of the magnetic field  in SXP1062. More direct methods -- especially cyclotron line measurements -- might be very important in this case.


Up to a very recent time the most popular models of magnetic field evolution in the case of magnetars were based on the properties of crustal fields. However, the magnetic field in the core can play an important role. 
Recently this question has been studied in \cite{2017arXiv170909167B}. Accounting for the field behavior in the core  might modify the standard set of assumptions about field evolution, especially in the case of magnetars. In particular, the time scales of field variations can become larger for some processes. If this is indeed the case, then other samples of sources might be used to test the hypothesis of the Hall attractor, especially for standard field values.  Recently, a new approach to describe field evolution in the core has been proposed \cite{2017arXiv170500508G}. Enhanced activity of theoreticians in analysing the behavior of magnetic fields in NS interiors can produce fruitful results, hopefully testable with observations.

Taking altogether, we conclude that different types of astrophysical sources can provide an opportunity to test predictions of magnetic field evolution scenarios and so to probe the existence of the Hall attractor. However, up to now there are no strong arguments in favor of this hypothesis. 

\section*{Acknowledgements}                    

SBP was supported by the grant from the Russian Science
Foundation 14-12-00146.

\section*{References}
\bibliographystyle{iopart-num}
\bibliography{hall}

\end{document}